# Observational Evidence of Pitch Angle Isotropization by IMF Waves

L. Saul, E. Möbius, C.W. Smith

University of New Hampshire, Durham NH 03824

P. Bochsler

Physikalisches Institut der Universität Bern, CH-3012 Bern

H. Grünwaldt

Max-Planck-Institut für Aeronomie, Katlenburg-Lindau

B. Klecker

Max-Planck-Institut für Extraterrestrische Physik, Garching

F. Ipavich

University of Maryland, College Park MD 20740

**Abstract.** A statistical analysis of interstellar $He^+$ pickup ion measurements from SOHO/CTOF combined with magnetic field data from WIND/MFI enable quantitative study of wave-particle interactions in the inner heliosphere for the first time. Magnetic field vector measurements with a time resolution of 3 seconds are used to determine power spectrum characteristics of interplanetary magnetic turbulence. These spectral characteristics are then compared in superposed epoch and correlation analyses with $He^+$ fluxes and spectra. The observed pickup ion velocity distributions can be explained consistently as a consequence of pitch angle scattering of the interstellar pickup ions by Alfvénic fluctuations.

## 1. Introduction

Magnetohydrodynamic waves play an important role in energetic particle and pickup ion transport. Alfvénic (transverse) fluctuations can act as pitch angle scatterers in the wave frame, which is important for particle transport processes. It has been assumed that diffusive transport and diffusive shock acceleration are dominated by magnetic field fluctuations [e.g., *Skilling*, 1971; *Lee* and *Völk*, 1975]. Quasi-linear theory uses the assumptions of small fluctuations and resonance to calculate pitch angle diffusion coefficients originating from different types of waves in the solar wind (SW) [e.g., *Schlickeiser*, 1998]. While theoretical descriptions of such wave-particle interactions abound, until now there has been no direct experimental verification of a correlation between the power in the fluctuations and the strength of observed transport effects. Because of their well defined initial distribution, pickup ions (PUIs) provide a tracer for studying such transport [*Chalov* and *Fahr*, 2002].

Neutral interstellar helium enters the heliosphere from the local interstellar medium and is ionized predominantly by solar ultraviolet radiation. The resulting PUIs are convected with the SW from their point of ionization, subject to such wave-particle interactions. Long before the first in situ detection of interstellar PUIs [*Möbius* et al, 1985], wave-particle effects had been suggested to influence energetic particles and PUIs [e.g., *Jokipii*, 1972 and references therein]. Detailed models of the evolution of these populations have been devised [*Vasilyunas* and *Siscoe*, 1976; *Isenberg*, 1997; *Schwadron*, 1998; *Zank* and *Pauls*, 1997] assuming pitch angle diffusion by magnetic turbulence. PUIs are injected (picked up) at a speed slow compared to the solar wind, and are created from a presumably constant interstellar neutral density, creating a ring distribution in velocity space. This well known, initially anisotropic, distribution of interstellar PUIs allows the study of their isotropization by pitch angle scattering, where other less well-known or more isotropic initial populations (e.g. cosmic rays) would not permit such quantitative analysis. In this paper, we compare PUI distributions with concurrent measurements of IMF wave power.

## 2. Instrumentation and Analysis

The CTOF instrument, in the CELIAS package on board SOHO, produced 150 days of PUI data under relatively steady conditions in the upwind region of the interstellar flow. The large geometric factor of the instrument makes high time resolution observations of PUI velocity spectra possible [*Hovestadt* et al., 1995]. Furthermore, its location at the Lagrangian point L1 and its 3-axis stabilized platform provides continuous sampling of the SW near 1AU. This makes the dataset interesting for studying the short term variations in both total flux and velocity distribution of interstellar $He^+$ PUIs. As SOHO is not equipped with a magnetometer, interplanetary magnetic field (IMF) measurements from the fluxgate magnetometers on WIND were used [*Lepping* et al., 1995]. During the time period considered (DOY 80-230, 1996) WIND passed through the Earth's magnetosphere 3 times, and was within $250R_E$ of SOHO. We ignored data from well before to well after the crossings to avoid terrestrial effects (DOY 86-89, 109-111, 130-133). The IMF measurements were shifted in time to correspond to SOHO's location, using the measured solar wind speed and IMF orientation, and assuming a magnetic field 'frozen' into the solar wind plasma. A direct comparison between SOHO/MTOF/PM and WIND/SWE solar wind parameters, as well as other spacecraft correlation studies [*Matsui* et al., 2002], suggest that this extrapolation technique is valid for the considered time resolution (>15 min.) and the parameters used in our study (wave amplitudes and IMF orientation). It should be noted however that due to the necessary convection of the data with the SW any correlation of the physical parameters may be reduced.

### 2.1. Wave Parameters

We used Fourier analysis in mean field coordinates (see below) to determine components of the power spectral density (PSD) of IMF variations. The frequency range considered is from 0.002 Hz. to 0.16 Hz (the Nyquist frequency for the 3s public WIND magnetometer data). It is not possible to uniquely identify frequency and wave vectors with only a single spacecraft, so this range of frequencies is used as an indicator of resonant wave power. The typical $He^+$ gyro-frequency in the SW is 0.015 Hz (2 nT) to 0.15 Hz (20 nT).

The PSD was computed for each contiguous 15 minute period in the dataset, and each component was fit to a power law with the least squares method. A sliding principal axis coordinate system was used to determine each PSD. The principal axes for each 3s field vector were calculated using a 15 minute sliding mean field direction $\hat{\bar{B}}$. The z component is chosen along the mean field, x along $\hat{\bar{B}} \times \hat{r}$, and y along $\hat{\bar{B}} \times (\hat{\bar{B}} \times \hat{r})$ [*Belcher* and *Davis*, 1971]. We consider here two components of the spectral matrix, the power in the z direc-tion $P_Z$, and the transverse power $P_{TR} = P(\sqrt{x^2 + y^2})$. $P_Z$ is the power in parallel (or compressive) fluctuations and $P_{TR}$ is the power in perpendicular (or transverse) fluctuations [*Matthaeus* and *Smith*, 1981]. The quantities $P_Z$ and $P_{TR}$ are taken from the power law fit, and are thus affected by wave power over the entire frequency range considered. The numerical value used for comparison is the value of the power law fit of the PSD component evaluated at 0.1 Hz.

### 2.2 PUI Parameters

PUIs are initially injected into the sunward hemisphere in velocity space and form a ring-like or toroidal velocity distribution, as identified by *Oka* et al. [2002]. Subsequently, PUIs pitch-angle scatter into a spherical shell, which shrinks due to adiabatic cooling in the expanding SW. Through continuous ionization PUIs fill a sphere in velocity space with radius $V_{SW}$. For near perpendicular IMF and/or rapid scattering this leads to a distribution with a sharp cutoff at $V = 2V_{SW}$ in the spacecraft frame [e.g., *Vasyliunas* and *Siscoe*, 1976]. If the IMF is nearly radial, injected PUIs must be pitch-angle scattered to enter the anti-sunward hemisphere. Therefore, a slow pitch-angle scattering rate leads to reduced PUI fluxes in this hemisphere during times of radial IMF, observed as a distinct anisotropy [*Gloeckler* et al., 1995]. This reduction is most prominent close to the cut-off as described in detail in *Möbius* et al. [1998]. Therefore, the energy flux near the cut-off, i.e. at $1.8 < V/V_{SW} < 2.0$ (in the spacecraft frame), which we call here for simplicity $He^+_C$, is a good proxy for the strength of PUI pitch angle scattering. This effect can be seen most clearly in comparison with distributions during times of perpendicular IMF, for which the velocity distributions are immediately isotropic.

The CTOF aperture points sunward, with an opening angle of 50°. Thus most of the PUI velocity distribution (convected anti-sunward with the solar wind and filling a sphere between 0 and 2 $V_{SW}$ in the spacecraft frame) is continuously in the field of view. However, the collecting power of electrostatic analyzers scales strongly with the ion energy, which reduces the detection efficiency of PUIs at the low energy end. In addition, it is difficult to separate PUIs from the solar wind at energies near that of the bulk SW. Therefore, we consider only PUIs with $V > V_{SW}$ in the spacecraft frame, or $V > 0$ in the SW frame, but this is the important part of the PUI distribution to observe the variation of flux with wave power during times of radial IMF, which should be most prominent close to the cut-off.

### 3. Observations

A sample time period is shown in Figure 1, with relatively calm solar wind conditions, a solar wind that was gradually decreasing from 600 to 450 km/s, and a magnetic field strength of 4 – 7 nT. The proton density is shown in the top panel, followed by the magnetic field orientation, the wave power, the PUI energy flux near the cut-off $He^+_C$, and the dynamic spectra. The helium PUI velocity distribution is depleted, especially near the cutoff ($He^+_C$ just above), when the IMF orientation changes to more radial, and high PUI fluxes in $He^+_C$ are seen for nearly perpendicular field as expected [e.g., *Möbius* et al., 1998]. However, the transverse wave power (3$^{rd}$ panel) is also lower by more than one order of magnitude during reduced PUI fluxes and thus may be correlated with $He^+_C$. However, it is difficult to see whether there are separate effects on the PUI distribution from variations in IMF orientation and/or the wave power. To distinguish between these effects, we pursue a statistical approach.

## 3.1 Superposed Epoch Analysis

To isolate the effects of wave power on the PUI distribution, superposed epoch analysis is applied to the full dataset. By combining PUI events from time periods with specified SW conditions and binning them in normalized velocity $V/V_{SW}$, spectra can be formed that are representative of the chosen SW conditions. For example, we can filter PUI events into those measured during times of near radial IMF and those measured during times of near perpendicular IMF and compare the resultant spectra.

The $He^+$ energy flux distributions for two ranges of IMF orientation, radial and perpendicular, are shown in Figure 2 and Figure 3, respectively. In each plot, the spectra are sorted into three ranges of transverse wave power, and the range of $He^+_C$ is indicated by dashed lines. These observations show a qualitative correlation of the PUI velocity distribution with $P_{TR}$, the transverse wave power. In both figures, the PUI flux in the range $1.6<V/V_{SW}<2.0$ increases with wave power. This somewhat unexpected result for perpendicular IMF (Fig. 3) could be due to correlations between PUI fluxes and proton density or field strength, particularly strong in compression regions [*Saul* et al., 2002], as wave power also correlates with these parameters (not shown here). To exclude the known strong effects from SW compression regions, only times with proton densities of $< 15$ cm$^{-3}$ were included. However, the density correlation is still visible. Because some weakening in the correlation of the wave phase was observed by Matsui et al. [2001] for separation distances >150 AU, we repeated the analysis restricted to smaller distances, which produced no significantly different result.

The most important result from Figures 2 and 3 is that the transverse wave power does not affect the shape of the cutoff for near perpendicular fields, whereas the steepness of the cutoff changes strongly with $P_{TR}$ in near radial fields. In particular, the cutoff during times of near radial IMF steepens with $P_{TR}$, towards the sharp cutoff present in near perpendicular IMF. This result is consistent with the hypothesis of pitch angle scattering by transverse waves, isotropizing the PUI distribution and creating a sharper cutoff even in regions of near radial IMF.

## 3.2 Statistical Correlations

To see whether this correlation between the wave power and the PUI distribution could be quantified, we use Pearson's correlation coefficient [e.g., *Richardson* and *Paularena*, 2001] to correlate the PUI and wave parameters. For two variables $X_i$ and $Y_i$, with n values that change over time with the index $i$ and whose averages are $\bar{X}$ and $\bar{Y}$, the correlation coefficient is defined as:

$$r = \frac{\sum_{i=0}^{n}(X_i - \bar{X})(Y_i - \bar{Y})}{\sqrt{\sum_{i=0}^{n}(X_i - \bar{X})^2 \cdot \sum_{j=0}^{n}(Y_j - \bar{Y})^2}}$$

We also want to determine the uncertainty of this coefficient. The correlation is not normally distributed, however its normally distributed transform ("Fisher's *z*") is within 1% of the correlation *r* for $r < 0.3$. For such correlations we take a statistical uncertainty equal to the standard deviation of that function [e.g., *Freund*, 1962]: $\Delta r \cong \sqrt{1/(n-3)}$.

Statistical correlations of PUI fluxes and wave power are shown in Table 1. As in Figures 2 and 3, only times with SW density <15 cm$^{-3}$ were included. The strongest correlation is observed between the energy flux of He$^+$ near the cutoff (He$^+_C$) and the transverse power in IMF fluctuations (P$_{TR}$). The range of He$^+_C$ is indicated with dashed lines in Figures 2 and 3. The correlation is calculated using all IMF orientations in Column 1. It increases when only periods of near radial field are considered (Column 3) and decreases for near perpendicular IMF (Column 2). This is again consistent with pitch-angle scattering by IMF waves isotropizing the PUIs in radial IMF. While the correlations in Table 1 are not large, they still support the more visible evidence in Figures 2 and 3.

## 4. Conclusions

We have shown the first observations of a connection between the intensity of waves and pitch-angle diffusion of particles. The observed correlation of PUI fluxes at the cutoff and IMF wave power during periods of radial IMF provides evidence for pitch angle scattering due to wave – particle interactions. While this correlation between waves and scattering has long enjoyed a strong theoretical backing, the observational evidence so far had been limited. This study gives both qualitative and quantitative evidence of PUI isotropization by transverse IMF waves.

However, this study does not yet provide direct insight or details of the scattering mechanism. In particular, the wave vectors have not been identified and so counter-propagation, resonant and/or non-resonant scattering, or other effects could be involved. The data set is also limited in time, being taken during solar minimum in relatively calm SW. It is possible that the observed correlation will change in SW with different flow and turbulent properties. Finally, the observed qualitative correlations with wave power at speeds closer to the bulk of the PUI velocity distribution (i.e. $1.6 < V/V_{SW} < 1.8$), that are even present in perpendicular IMF, have no explanation at this time.

**Acknowledgments.** This work is partially supported by NASA grants NGT5-50381, NAG5-10890 and the Swiss National Foundation. We are grateful to Phil Isenberg, Marty Lee, and Hiroshi Matsui for helpful discussions, and to Yuri Litvinenko for assistance with the data sets. The WIND MFI team is gratefully acknowledged for the magnetic field data.

## References

Belcher, J.W., and L.J. Davis, Large Amplitude Alfven Waves in the Interplanetary Medium II, *Journal of Geophysical Research*, *76*, 3534, 1971.

Chalov, S.V., and H.J. Fahr, Different solar wind types reflected in pick-up ion spectral signatures, *Astron. Astrophys.*, *384*, 299-302, 2002.

Freund, J.E., *Mathematical Statistics*, Prentice-Hall, Englewood Cliffs, NJ, 1962.

Gloeckler, G., N.A. Schwadron, L.A. Fisk, and J. Geiss, Weak Pitch Angle Scattering of few MV Rigidity Ions from Measurements of Anisotropies in the Distribution Function of Interstellar Pickup H+, *Geophysical Research Letters*, *22* (19), 2665-2668, 1995.

Hovestadt, D. et al., CELIAS: The Charge Element and Isotope Analysis System for SOHO, *Solar Physics*, *162*, 441-481, 1995.

Isenberg, P.A., A hemispherical model of anisotropic interstellar pickup ions, *Journal of Geophysical Research*, *102* (A3), 4719-4724, 1997.

Jokipii, J.R., Fokker-planck equations for charged-particle transport in random fields, *Astrophysical Journal*, *172*, 319-326, 1972.

Lee, M.A., and H.J. Völk, Hydromagnetic Waves and Cosmic-Ray Diffusion Theory, *Astrophysical Journal*, *198*, 485-492, 1975.

Lepping, R.P., et al., The WIND Magnetic Field Investigation, *Space Science Reviews*, *71*, 207, 1995.

Matsui, H., C.J. Farrugia, and R.B. Torbert, Wind-ACE soalr wind correlations, 1999: An approach through spectral analysis, *Journal of Geophysical Research*, *107* (A11), 1355, 2002.

Matthaeus, W.H., and C. Smith, Structure of correlation tensors in homogeneous anisotropic turbulence, *Physical Review A*, *24* (4), 2135-2144, 1981.
Möbius, E., D. Hovestadt, B. Klecker, M. Scholer, G. Gloeckler, and F.M. Ipavich, Direct observation of He+ pick-up ions of interstellar origin in the solar wind, *Nature*, 1985.
Möbius, E., D. Rucinski, M.A. Lee, and P.A. Isenberg, Decreases in the antisunward flux of interstellar pickup He+ associated with radial interplanetary magnetic field, *Journal of Geophysical Research*, *103*, 257, 1998.
Oka, M., T. Terasawa, H. Noda, Y. Saito, and T. Mukai, 'Torus' distribution of interstellar pickup ions: Direct observation, *Geophysical Research Letters*, *29* (12), 1612, 2002.
Richardson, J.D., and K.I. Paularena, Plasma and Magnetic Field Correlations in the Solar Wind, *Journal of Geophysical Research*, *106* (A1), 239-251, 2001.
Saul, L., E. Möbius, Y. Litvinenko, P.A. Isenberg, H. Kucharek, M.A. Lee, H. Grünwaldt, F.M. Ipavich, B. Klecker, and P. Bochsler, SOHO CTOF Observations of Interstellar He+ Pickup Ion Enhancements in Solar Wind Compression Regions, in *Tenth International Solar Wind Conference*, edited by M. Velli, R. Bruno, and F. Malara, pp. 778, AIP, Pisa, Italy, 2002.
Schlickeiser, R., Quasi-linear theory of cosmic ray transport and acceleration: the role of oblique magnetohydrodynamic waves and transit-time damping, *Astrophysical Journal*, *492*, 352-378, 1998.
Schwadron, N.A., A model for pickup ion transport in the heliosphere in the limit of uniform hemispheric distributions, *Journal of Geophysical Research*, *103* (A9), 20643-20649, 1998.
Skilling, J., Cosmic Rays in the Galaxy: Convection or Diffusion?, *Astrophysical Journal*, *170*, 265-273, 1971.
Vasyliunas, V.M., and G.L. Siscoe, On the flux and energy spectrum of interstellar ions in the solar system, *Journal of Geophysical Research*, *81*, 1247-1252, 1976.
Zank, G.P., and H.L. Pauls, Shock propagation in the outer heliosphere 1. Pickup ions and gasdynamics, *Journal of Geophysical Research*, *102* (A4), 7037-7050, 1997.

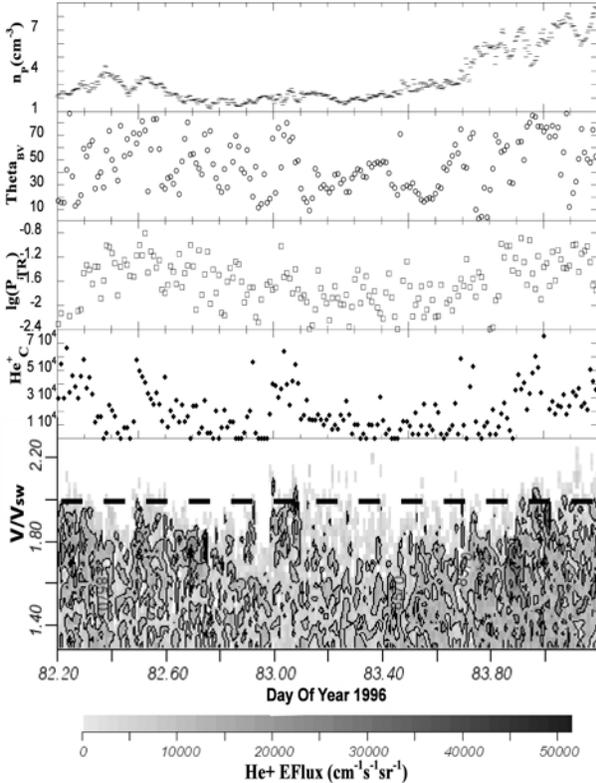

**Figure 1.** PUI spectra and related SW data for DOY 82.2 – 84.2, 1996. Top panel: proton density from SOHO/MTOF/PM. 2nd panel: IMF angle from radial. 3rd panel: $\log_{10}$ of transverse wave power ($B^2$). 4th panel: energy flux density of He$^+$ near the cut-off velocity (see text). Bottom panel: Dynamic PUI energy flux spectra from SOHO/CTOF, with the normalized velocity on the vertical axis. The expected cut-off at $V/V_{SW} = 2$ is shown with a dashed line.

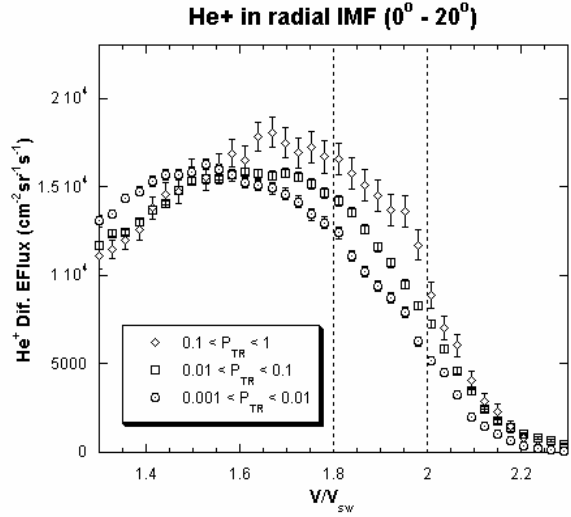

**Figure 2.** PUI energy flux spectra for three ranges of $P_{TR}$, or transverse wave power, restricted to near parallel IMF fields. Time periods with proton densities >15cm$^{-3}$ were excluded. Error bars represent the statistical error.

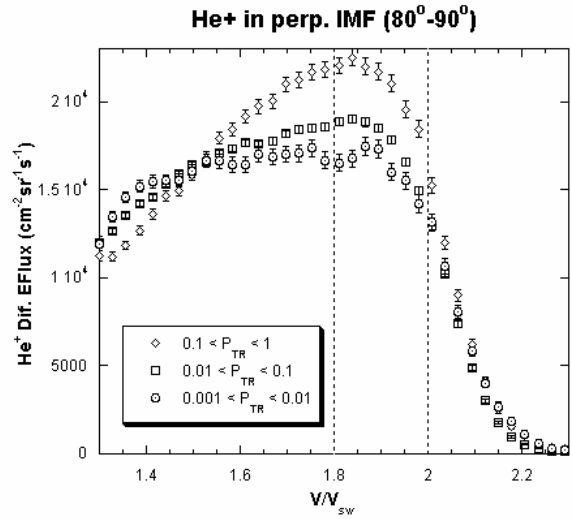

**Figure 3.** PUI energy flux spectra for three ranges of $P_{TR}$, restricted to near perpendicular IMF fields (similar to Fig. 2).

**Table 1.** Wave/PUI parameter correlation coefficients

| ($r$) | $He^{+}_{C}$ | $He^{+}_{C-(Perp.)}$ | $He^{+}_{C-(Par.)}$ |
|---|---|---|---|
| $P_{TR}$ | 0.19 ± .01 | 0.12 ± .03 | 0.21 ± .06 |
| $P_{Z}$ | 0.14 ± .01 | 0.00 ± .03 | 0.19 ± .06 |

Second column includes only IMF within 80° - 90° from radial; the last column within 0° - 10° from radial.